\documentclass[letterpaper,preprintnumbers,prd,nofootinbib,nobibnotes,showpacs]{revtex4}
\usepackage{amsfonts}
\usepackage{mathrsfs}
\usepackage{epsfig}
\usepackage{graphicx}%
\usepackage{dcolumn}
\usepackage{amsmath}

\makeatletter
\def\btt#1{\texttt{\@backslashchar#1}}%
\DeclareRobustCommand\bblash{\btt{\@backslashchar}}%
\makeatother
\begin{document}

\title{A Static Spherically Symmetric Solution of the Einstein-aether Theory}
\author{Changjun Gao}\email{gaocj@bao.ac.cn} \affiliation{The National
Astronomical Observatories, Chinese Academy of  Sciences, Beijing,
100012, China} \affiliation{{$^{}$Kavli Institute for Theoretical
Physics China, CAS, Beijing 100190, China }}
\author{You-Gen Shen}
\email{ygshen@center.shao.ac.cn} \affiliation{Shanghai
Astronomical Observatory, Chinese Academy of Sciences, Shanghai
200030, China}

\date{\today}

%%%%%%%%%%%%%%%%%%%%%%%%%%%%%%%%%%%%%%%%%%%%%%%%%%%%%%%%%%%%%%%%%%%%%%%%%%
\begin{abstract}
By using of the Euler-Lagrange equations, we find a static
spherically symmetric solution in the Einstein-aether theory with
the coupling constants restricted. The solution is similar to the
Reissner-Nordstrom solution in that it has an inner Cauchy horizon
and an outer black hole event horizon. But a remarkable difference
from the Reissner-Nordstrom solution is that it is not
asymptotically flat but approaches a two dimensional sphere. The
resulting electric potential is regular in the whole spacetime
except for the curvature singularity. On the other hand, the
magnetic potential is divergent on both Cauchy horizon and the
outer event horizon.
\end{abstract}

%\keywords{}

\maketitle

%%%%%%%%%%%%%%%%%%%%%%%%%%%%%%%%%%%%%%%%%%%%%%%%%%%%%%%%%%%%%%%%%%%%%%%%%%
\section{Introduction}
The Einstein-aether theory \cite {ted:00,ted:07} belongs to the
vector-tensor theories in nature. Besides the ordinary matters and
the metric tensor $g_{\mu\nu}$, the fundamental field in the
theory is a timelike vector field $A_{\mu}$. Different from the
usual vector-tensor theories, $A_{\mu}$ is constrained to have a
constant norm. So the vector field $A_{\mu}$ cannot vanish
anywhere. Therefore, a preferred frame is defined and the Lorentz
symmetry is violated. The vector field is referred to as the
``aether''. The Einstein-aether theory has become an interesting
theoretical laboratory to explore both the Lorentz violation
effects and the preferred frame effects. Up to now, the
Einstein-aether theory has been widely studied in literature in
various ways: the analysis of classical and quantum perturbations
\cite{lim:05,na:10,car:00,car:01,car:02,car:03}, the cosmologies
\cite{car:04,bon:08}, the gravitational collapse \cite{gar:08},
the Einstein-aether waves \cite{wave:04}, the radiation damping
\cite{fos:06} and so on.

The purpose of the present paper is to seek for a static
spherically symmetric solution of the Einstein-aether theory. The
black hole solutions in the Einstein-aether theory have been
investigated in Refs. \cite{bh1,bh2,bh3,bh4,bh5}. These
investigations mainly focus on the numerical analysis of the
solutions due to the complication of the Einstein equations. To
our knowledge, one have not yet find the exact, static and
spherically symmetric solution in the Einstein-aether theory. In
this paper, instead of solving the Einstein equations, we are
going to solve the Euler-Lagrange equations in order to derive the
static spherically symmetric solution. We find it is relatively
simple in the calculations. We shall use the system of units in
which $16\pi G=c=\hbar=4\pi\varepsilon_0=1$ and the metric
signature $(-,\ +,\ +,\ +)$ throughout the paper.

\section{static spherically symmetric solution}
In the context of spherical symmetry and after the redefinitions
of metric $g_{\mu\nu}$ and aether field $A_{\mu}$, the Lagrangian
density of the Einstein-aether theory can be written as
\begin{eqnarray}
\label{eq:lagM}
\mathscr{L}&=&-R-\frac{c_1}{2}F_{\mu\nu}F^{\mu\nu}-c_2\left(\nabla_{\mu}A^{\mu}\right)^2
+\lambda\left(A_{\mu}A^{\mu}+m^2\right)\;,
\end{eqnarray}
with the field strength tensor
\begin{eqnarray}
F_{\mu\nu}&=&\nabla_{\mu}A_{\nu}-\nabla_{\nu}A_{\mu}\;.
\end{eqnarray}
Here $R$ is the Ricci scalar and the $c_i$ are dimensionless
constants. We note that there is a sign difference from
\cite{gar:08} in the definition of Ricci tensor. $\lambda$ is the
Lagrange multiplier field which has the dimension of the square of
inverse length, $l^{-2}$. $m$ is a positive dimensionless constant
which has the physical meaning of the squared norm for the aether
field. The requirement of $m^2>0$ ensures the aether to be
timelike.

The static and spherically symmetric metric can always be written
as
\begin{eqnarray}\label{eq:line}
ds^2=-U\left(r\right)dt^2+\frac{1}{U\left(r\right)}dr^2+f\left(r\right)^2d\Omega^2\;.
\end{eqnarray}
Instead of solving the Einstein equations, we prefer to deal with
the Euler-Lagrange equations from the Lagrangian
Eq.~(\ref{eq:lagM}) for simplicity in calculations. Because of the
static and spherically symmetric property of the spacetime, the
vector field $A_{\mu}$ takes the form
\begin{eqnarray}\label{eq:A}
A_{\mu}=\left[\phi\left(r\right),\ \frac{1}{\psi\left(r\right)},\
0,\ 0\right]\;,
\end{eqnarray}
where $\phi$ and $\psi^{-1}$ correspond to the electric and
magnetic part of the electromagnetic potential.  Then we have
\begin{eqnarray}
F_{\mu\nu}F^{\mu\nu}=-2\phi^{'2}\;,\ \ \
\nabla_{\mu}A^{\mu}=\left(\frac{U}{\psi}\right)^{'}+2\frac{f^{'}U}{f\psi}\;.
\end{eqnarray}
The prime here and in what follows denotes the derivative with
respect to $r$. Taking into account the Ricci scalar, $R$, we have
the total Lagrangian as follows

\begin{eqnarray}
\label{eq:lagMM}
\mathscr{L}&=&-U^{''}-4U^{'}\frac{f^{'}}{f}-4U\frac{f^{''}}{f}+\frac{2}{f^2}-2U\frac{f^{'2}}{f^2}
\nonumber\\&&+{c_1}\phi^{'2}-c_2\left[\left(\frac{U}{\psi}\right)^{'}+2\frac{f^{'}U}{f\psi}\right]^2
\nonumber\\&&+\lambda\left(-\frac{1}{U}\phi^2+\frac{U}{\psi^2}+m^2\right)\;.
\end{eqnarray}
Let
\begin{eqnarray}
\psi=\frac{Uf^2}{K}\;,
\end{eqnarray}
we can rewrite the Lagrangian, Eq.~(\ref{eq:lagMM}), as follows

\begin{eqnarray}
\label{eq:lagMMM}
\mathscr{L}&=&-U^{''}-4U^{'}\frac{f^{'}}{f}-4U\frac{f^{''}}{f}+\frac{2}{f^2}-2U\frac{f^{'2}}{f^2}
\nonumber\\&&+{c_1}\phi^{'2}-c_2\frac{K^{'2}}{f^4}
\nonumber\\&&+\lambda\left(-\frac{1}{U}\phi^2+\frac{K^2}{Uf^4}+m^2\right)\;.
\end{eqnarray}
Now there are $U,\ f,\ \phi,\ K,\ \lambda$ five variables in the
Lagrangian which correspond to five equations of motion. Then
using the Euler-Lagrange equation, we obtain the equation of
motion for $\lambda$,

\begin{eqnarray}\label{eq:lambda}
-\frac{1}{U}\phi^2+\frac{K^2}{Uf^4}+m^2=0\;,
\end{eqnarray}
for $\phi$,

\begin{eqnarray}
\label{eq:phi} c_1Uf\phi^{''}+2c_1U\phi^{'}f^{'}+\lambda\phi
f=0\;,
\end{eqnarray}
for $K$,

\begin{eqnarray}\label{eq:K}
c_2UfK^{''}-2c_2U K^{'}f^{'}+\lambda K f=0\;,
\end{eqnarray}
for $U$,

\begin{eqnarray}\label{eq:U}
-2U^2f^3f^{''}-\lambda  K^{2}+\lambda \phi^2 f^4=0\;,
\end{eqnarray}
and for$f$,
\begin{eqnarray}\label{eq:f}
&&-c_1\phi^{'2}Uf^4-c_2UK^{'2}+\lambda\phi^2f^4+Uf^4U^{''}-\lambda
m^2 Uf^4\nonumber\\&&+2Uf^3U^{'}f^{'}+2U^2f^3f^{''}+\lambda
K^2=0\;,
\end{eqnarray}
respectively. We have five independent differential equations and
five variables, $U,\ f,\ \phi,\ K,\ \lambda$. So the system of
equations is closed.

From Eq.~(\ref{eq:lambda}) and Eq.~(\ref{eq:phi}), we obtain
\begin{eqnarray}\label{eq:sU}
U=\frac{\phi^2f^4-K^2}{m^2f^4}\;,
\end{eqnarray}
and
\begin{eqnarray}\label{eq:sphi}
\lambda=-\frac{c_1U\left(f\phi^{''}+2\phi^{'}f^{'}\right)}{f\phi}\;,
\end{eqnarray}
respectively. Substituted Eq.~(\ref{eq:sU}) and
Eq.~(\ref{eq:sphi}) into Eq.~(\ref{eq:U}), then Eq.~(\ref{eq:U})
becomes

\begin{eqnarray}\label{eq:ssU}
2\phi f^{''}+c_1m^2f\phi^{''}+2c_1m^2\phi^{'}f^{'}=0\;.
\end{eqnarray}

The norm of the aether field is usually constrained by the
Lagrange multiplier to be unity, $m=1$. But in this paper, we
would constrain the norm to meet

\begin{eqnarray}\label{eq:norm}
c_1m^2=2\;.
\end{eqnarray}
We note that this choice is consistent with the perturbation
analysis of Lim \cite{lim:05}. He showed that in order to have a
positive definite Hamiltonian, $c_1$ should satisfy
\begin{eqnarray}\label{eq:lim}
c_1>0\;.
\end{eqnarray}
We stress that the choice of $c_1m^2=2$ corresponds to the special
case that has been called $c_{14}=2$ \cite{eling04,car:04} which
leads the Newton's gravitational constant to
infinity\cite{eling04,car:04}:

\begin{eqnarray}\label{eq:G}
G_{N}=\frac{G}{1-c_{14}/2}\;.
\end{eqnarray}
But Eq.~(\ref{eq:G}) should be taken with a grain of salt because
it is derived with vanishing spatial components in $A^{\mu}$. We
also stress that the choice of $c_1m^2=2$ is not `` for
convenience '' but actually a restriction on the theory
\footnote{We thank Ted Jacobson for bringing these points to our
notice.}.

Then Eq.~(\ref{eq:ssU}) gives the solution as follows
\begin{eqnarray}\label{eq:ssphi}
\phi=\frac{\phi_0+\phi_1 r}{f}\;,
\end{eqnarray}
where $\phi_0,\ \phi_1$ are two integration constants. $\phi_1$ is
dimensionless while $\phi_0$ has the dimension of length.

Keeping Eqs.~(\ref{eq:sU}), (\ref{eq:sphi}) and (\ref{eq:ssphi})
in mind, we find Eq.(\ref{eq:K}) and Eq.~(\ref{eq:f}) are reduced
to the following form

\begin{eqnarray}\label{eq:33}
2c_2m^2 K^{'}f^{'}-c_2m^2f K^{''}-2K f^{''}=0\;,
\end{eqnarray}
and
\begin{eqnarray}\label{eq:55}
&&2f^2K^{'2}+12K^2 f^{'2}-12fK f^{'}K^{'}+c_2m^2f^2
K^{'2}\nonumber\\&&-6fK^2 f^{''}+2Kf^2 K^{''}=0\;,
\end{eqnarray}
respectively. Putting

\begin{eqnarray}\label{eq:alpha}
&&c_2=\frac{1}{\alpha m^2}\;,
\end{eqnarray}
with $\alpha$ a new dimensionless parameter, we obtain from
Eq.~(\ref{eq:33}) and Eq.~(\ref{eq:55})

\begin{eqnarray}\label{eq:eqa}
&&6\alpha f K^2 f^{''}-12\alpha K^2 f^{'2}+8\alpha f K
f^{'}K^{'}\nonumber\\&&+4\alpha^2f K^2 f^{''}-2\alpha f^2
K^{'2}-f^2 K^{'2}=0\;.
\end{eqnarray}
In order that the spin-$0$ field does not propagate
superluminally, Lim constrained $c_2$ to meet \cite{lim:05}

\begin{eqnarray}
c_2>0\;, \ \ \ \ \textrm{and}\ \ \ \frac{c_2}{c_1}\leq 1\;.
\end{eqnarray}
Taking account of Eq.~(\ref{eq:lim}), we conclude that $\alpha$
should satisfy

\begin{eqnarray}
\alpha\geq\frac{1}{2}\;.
\end{eqnarray}

Solving the differential equation, we obtain

\begin{eqnarray}\label{eq:KKK}
&&K=K_0\exp\left\{{\int\frac{1}{\left(1+2\alpha\right)f}\left[4\alpha
f^{'}+\sqrt{2\alpha\left(2\alpha+3\right)\left(2\alpha f f^{''}+f
f^{''}-2f^{'2}\right)}\right]dr}\right\}\;,
\end{eqnarray}
where $K_0$ is an integration constant which has the dimension of
the square of the length, $l^{2}$. We may assume $K_0>0$.
Substituting Eq.~(\ref{eq:KKK}) into Eq.~(\ref{eq:55}), we obtain

\begin{eqnarray}\label{eq:chang}
&&\left(16\alpha f f^{''}+8 f f^{''}-16
f^{'2}\right)\sqrt{\alpha\left(2\alpha+3\right)\left(2\alpha f
f^{''}+ff^{''}-2f^{'2}\right)}\nonumber\\&&+\sqrt{2}\left(4\alpha^2f^2f^{'''}+12\alpha^2ff^{'}f^{''}-8\alpha
f^{'3}+4\alpha f^2 f^{'''}-12\alpha
ff^{'}f^{''}-9ff^{'}f^{''}+f^2f^{'''}+12f^{'3}\right)=0\;.
\end{eqnarray}
At first glance, Eq.~(\ref{eq:chang}) is rather complicated. But
using the calling sequence of ``dsolve'' in Maple Program, it is
easy to find the solutions with $\alpha=1/2,\ \ \alpha=3/2, \ \
\alpha=5/2,\ \cdot\cdot\cdot$. For general $\alpha$, $f$ is found
to be

\begin{eqnarray}\label{eq:fs}
&&f=f_0\left(1-k^2r^2\right)^{\frac{2\alpha+1}{4\alpha-2}}e^{\frac{\sqrt{6\alpha+4\alpha^2}}{2\alpha-1}\tanh^{-1}k
r}\;,
\end{eqnarray}
where $f_0$ and $k$ are integration constants. Both $f_0$ and
$k^{-1}$ have the dimension of length. Without the loss of
generality, the third integration constant with the dimension of
length has been absorbed by $r$.

Eq.~(\ref{eq:fs}) forces $k r$ to satisfy
\begin{eqnarray}
-1\leq k r\leq 1\;.
\end{eqnarray}
Up to this point, we could present all the variables:

\begin{eqnarray}\label{eq:all}
f&=&f_0\left(1-k^2r^2\right)^{\frac{2\alpha+1}{4\alpha-2}}e^{\frac{\sqrt{6\alpha+4\alpha^2}}{2\alpha-1}\tanh^{-1}k
r}\;,\\
 K&=&K_0\left(1-k
r\right)^{\frac{4\alpha-\sqrt{2\alpha\left(2\alpha+3\right)}}{4\alpha-2}}\cdot\left(1+k
r\right)^{\frac{4\alpha+\sqrt{2\alpha\left(2\alpha+3\right)}}{4\alpha-2}}\;,\\
\phi&=&\frac{1}{f_0}\left(\phi_0+\phi_1
r\right)\left(1-k^2r^2\right)^{-\frac{2\alpha+1}{4\alpha-2}}e^{-\frac{\sqrt{6\alpha+4\alpha^2}}{2\alpha-1}\tanh^{-1}k
r}\;,\\
U&=&\frac{1}{m^2f_0^2}\left(\phi_0+\phi_1
r\right)^2\left(1-k^2r^2\right)^{-\frac{2\alpha+1}{2\alpha-1}}e^{-\frac{2\sqrt{6\alpha+4\alpha^2}\tanh^{-1}k
r}{2\alpha-1}}\nonumber\\&&-\frac{K_0^2}{m^2f_0^4}\left(1-k
r\right)^{\frac{4\alpha-\sqrt{2\alpha\left(2\alpha+3\right)}}{2\alpha-1}}\cdot\left(1+k
r\right)^{\frac{4\alpha+\sqrt{2\alpha\left(2\alpha+3\right)}}{2\alpha-1}}\left(1-k^2r^2\right)^{-\frac{4\alpha+2}{2\alpha-1}}
e^{-\frac{4\sqrt{6\alpha+4\alpha^2}\tanh^{-1}k
r}{2\alpha-1}}\;,\\
\psi&=&\frac{\left(\phi_0+\phi_1r\right)^2\left(1-k^2r^2\right)^{-\frac{4\alpha+\sqrt{6\alpha
+4\alpha^2}}{4\alpha-2}}\left(1-kr\right)^{\frac{\sqrt{6\alpha+4\alpha^2}}{2\alpha-1}}}{K_0m^2}\nonumber\\&&-\frac{K_0\left(1-k
r\right)^{\frac{4\alpha}{2\alpha-1}}\left(1+k
r\right)^{\frac{4\alpha+\sqrt{6\alpha+4\alpha^2}}{2\alpha-1}}}
{m^2f_0^2\left(1-k^2r^2\right)^{\frac{2+8\alpha+\sqrt{6\alpha+4\alpha^2}}{4\alpha-2}}e^{\frac{2\sqrt{6\alpha+4\alpha^2}\tanh^{-1}k
r}{2\alpha-1}}}\;,\\
\lambda&=&-\frac{2k^2\left(\phi_0+\phi_1 r\right)^2}
{f_0^2m^4\left(2\alpha-1\right)^2\left(1-k^2r^2\right)^{\frac{6\alpha-1}{2\alpha-1}}}\left[4k
r\sqrt{4\alpha^2+6\alpha}-6\alpha-2k^2r^2-1-4\alpha
k^2r^2\right]e^{-\frac{2\sqrt{6\alpha+4\alpha^2}\tanh^{-1}k
r}{2\alpha-1}}\nonumber\\&&+\frac{2k^2K_0^2}
{f_0^4m^4\left(2\alpha-1\right)^2\left(1-k^2r^2\right)^{\frac{6\alpha-1}{2\alpha-1}}}\left[4k
r\sqrt{4\alpha^2+6\alpha}-6\alpha-2k^2r^2-1-4\alpha
k^2r^2\right]\nonumber\\&&\cdot\left(1-k
r\right)^{\frac{4\alpha-\sqrt{2\alpha\left(2\alpha+3\right)}}{2\alpha-1}}\cdot\left(1+k
r\right)^{\frac{4\alpha+\sqrt{2\alpha\left(2\alpha+3\right)}}{2\alpha-1}}e^{-\frac{4\sqrt{6\alpha+4\alpha^2}\tanh^{-1}k
r}{2\alpha-1}}\;.
\end{eqnarray}
If we define
\begin{eqnarray}
\phi_0\equiv\alpha_0f_0\;,\ \ \ \phi_1\equiv\alpha_1\;,\ \ \
k\equiv \alpha_2\frac{1}{f_0}\;,\ \ \ K_0\equiv\alpha_3 f_0^2\;,
\end{eqnarray}
then $\alpha_i$ are dimensionless constants. Together with $m$ and
$\alpha$, we have totally six dimensionless constants and one
dimensional parameter, $f_0$. We note that the seven parameters
are not independent and there are six parameters in the solution
in nature. In fact, Eling and Jacobson ~\cite{bh3} have argued
that there is a $3$-parameter (corresponding to the mass, electric
charge and magnetic charge, respectively) family of spherical,
static solutions before asymptotic flatness and regularity are
imposed. If one take into account the two coupling constants, that
would be $5$ parameters in all. But our solution Eqs.~(31-36) is
not asymptotic flat. So there is an extra parameter of
``cosmological-constant-like''. Then the total number of
parameters is six. \footnote{We thank Ted Jacobson for pointing
out this point.} This could be understood from the expression of
$f$ and $U$ with the replacements

\begin{eqnarray}
\frac{\phi_0}{mf_0}\rightarrow\bar{\phi_0}\;,\ \ \
\frac{\phi_1}{mf_0}\rightarrow\bar{\phi_1}\;,\frac{K_0}{mf_0^2}\rightarrow\bar{K_0}\;.
\end{eqnarray}
Then the metric of spacetime is determined by six parameters.
\section{structure of the spacetime }
In this section, let's numerically study the structure of
spacetime described by the solution. Since $f$ is the physical
length, we should rewrite the metric as follows

\begin{eqnarray}
ds^2=-U\left(r\right)dt^2+\frac{1}{V\left(r\right)}df^2+f\left(r\right)^2d\Omega^2\;,
\end{eqnarray}
with
\begin{eqnarray}
V\left(r\right)={U\left(r\right)f^{'2}}\;.
\end{eqnarray}
Now $f$ plays the role of physical radius  (proper length) of the
static spherically symmetric space.

As an example, we put the dimensional constant $f_0=1$(for
example, $f_0$ equals to one Schwarzschild radius). Five
dimensionless constants are put $m=1,\ \alpha_0=\alpha_2=1,\ \
\alpha=3/2$. As for $\alpha_1$, we let $\alpha_1=0.15,\ 0.1,\ 0,\
-0.2,\ -0.4$, respectively.

There are usually two kinds of horizons in a static spherically
symmetric spacetime, namely, the timelike limit surface (TLS) and
the event horizon (EH). The timelike limit surface separates the
timelike region of the Killing vector field from the spacelike
part which is determined by \cite{Haw73}
\begin{eqnarray}
g_{00}=U=0\;.
\end{eqnarray}

\begin{figure}
\includegraphics[width=8.4cm]{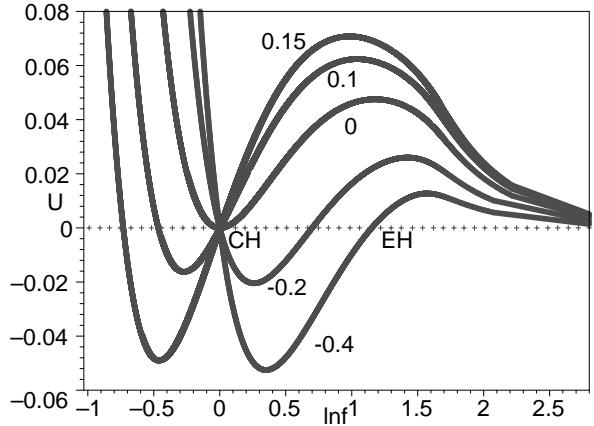}
\\
\caption{The evolution of $U$ with respect to the physical radius
$\ln{f}$ for different $\alpha_1=-0.4,\ -0.2,\ 0, \, 0.1,\ 0.15$.
There are two horizons in the spacetime in general, the inner
Cauchy horizon ($\textrm{CH}$) and the black hole event horizon
($\textrm{EH}$) (As an example, the $\textrm{CH}$ and
$\textrm{EH}$ are given for $\alpha_1=-0.4$). When $f\rightarrow
\infty$, we have $U=0$. So the solution is not asymptotically flat
in space. With the increasing of $\alpha_1$, the event horizon is
shrinking. When $\alpha_1=0$, the inner Cauchy horizon and the
black hole event horizon coincide and the solution corresponds to
the extreme solution.} \label{fig:fu}
\end{figure}

In Fig.~\ref{fig:fu}, we plot the evolution of $U$ with respect to
the physical radius $\ln f$ for different $\alpha_1$. The figure
shows that there are two TLS in the spacetime in general. One of
them is the inner Cauchy horizon and the other is the black hole
event horizon. This is very similar to the spacetime of
Reissner-Nordstrom solution. \footnote{A spacetime is globally
hyperbolic if there exist Cauchy surfaces (not Cauchy horizon) in
the spacetime. In the Reissner-Nordstrom (RN) spacetime, the
timelike property of the curvature singularity reveals it is not
globally hyperbolic. So there is no Cauchy surface in the RN
spacetime. But there is an inner Cauchy horizon in the RN
spacetime \cite{wald84}. Similarly, the singularity in our
solution is also timelike because of $g_{00}<0$ and $g_{11}>0$
when $0<f<f_{CH}$ ($f_{CH}$ represents the radius of Cauchy
horizon) and our solution is not globally hyperbolic. When
$f_{CH}<f<f_{EH}$ ($f_{EH}$ represents the radius of black  hole
event horizon), we have $g_{00}>0$ and $g_{11}<0$. It is a
spacelike region. Furthermore, there exists a curvature
singularity within the event horizon. So the spacetime is for a
black hole. When $f>f_{EH}$, we have $g_{00}<0$ and $g_{11}>0$
which is again a timelike region.} On the other hand, the
Reissner-Nordstrom spacetime is asymptotically flat in space. But
this solution is asymptotically a two dimensional sphere.
\footnote{When $f\rightarrow \infty$, we find $g_{00}=0$ and
$g_{11}=0$. The metric becomes $ds^2=f^2d\Omega^2$ which is for a
two dimensional sphere.} With the increasing of $\alpha_1$, the
event horizon is shrinking. When $\alpha_1=0$, the inner Cauchy
horizon and the black hole event horizon coincide and the solution
corresponds to the extreme solution.

\begin{figure}
\includegraphics[width=8.4cm]{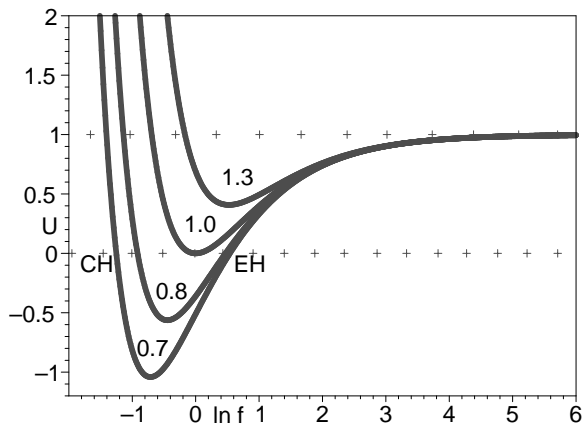}
\\
\caption{The evolution of $U$ with respect to the physical radius
${f}$ in the Reissner-Nordstrom solution for different electric
charge $Q=1.3,\ 1.0,\ 0.8,\ 0.7$. There are two horizons in the
spacetime, the inner Cauchy horizon ($\textrm{CH}$) and the black
hole event horizon ($\textrm{EH}$). (As an example, the
$\textrm{CH}$ and $\textrm{EH}$ are given for $Q=0.7$). The space
is asymptotically flat. With the increasing of electric charge
$Q$, the EH is shrinking and the CH expanding. When $Q=M=1.0$, the
inner Cauchy horizon and the black hole event horizon coincide and
the solution corresponds to the extreme solution.} \label{fig:RND}
\end{figure}

Compared to Fig.~\ref{fig:fu}, the structure of Reissner-Nordstrom
spacetime is shown in Fig.~\ref{fig:RND}. The metric of
Reissner-Nordstrom spacetime takes the form of

\begin{eqnarray}
ds^2&=&-Udt^2+\frac{1}{U}df^2+f^2d\Omega^2\;,\ \ \ \ \nonumber\\
U&=&1-\frac{2M}{f}+\frac{Q^2}{f^2}\;.
\end{eqnarray}
Without the loss of generality, we take the mass $M=1.0$ and the
electric charge $Q=1.3,\ 1.0,\ 0.8,\ 0.7$, respectively. There are
two horizons in the spacetime, the inner Cauchy horizon
($\textrm{CH}$) and the black hole event horizon ($\textrm{EH}$).
(As an example, the $\textrm{CH}$ and $\textrm{EH}$ are given for
$Q=0.7$). The space is asymptotically flat. With the increasing of
electric charge $Q$, the EH is shrinking and the CH expanding.
When $Q=M=1.0$, the inner Cauchy horizon and the black hole event
horizon coincide and the solution corresponds to the extreme
solution.

\begin{figure}
\includegraphics[width=8.4cm]{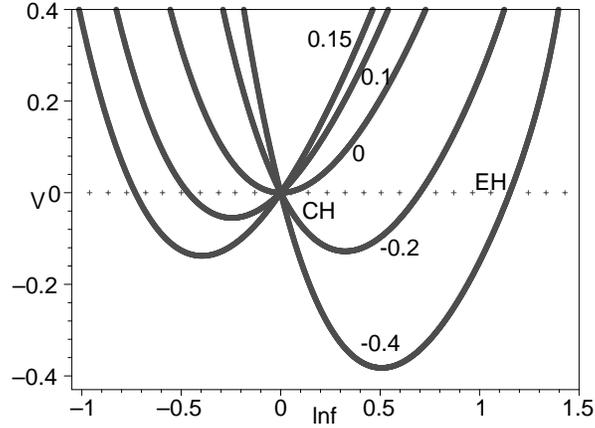}
\\
\caption{The evolution of $V$ with respect to the physical radius
$\ln{f}$ for different $\alpha_1=-0.4,\ -0.2,\ 0, \, 0.1,\ 0.15$.
There are two horizons in the spacetime in general, the inner
Cauchy horizon and the black hole event horizon (As an example,
the $\textrm{CH}$ and $\textrm{EH}$ are given for
$\alpha_1=-0.4$). With the increasing of $\alpha_1$, the event
horizon is shrinking. When $\alpha_1=0$, the inner Cauchy horizon
and the black hole event horizon coincide and the solution
corresponds to the extreme solution.} \label{fig:fv}
\end{figure}

On the other hand, the EH is determined by \cite{Haw73}

\begin{eqnarray}
g^{11}=V=0\;.
\end{eqnarray}
In Fig.~\ref{fig:fv}, we plot the evolution of $V$ with respect to
the physical radius $\ln f$ for different $\alpha_1$. The figure
shows that there are two horizons in the spacetime in general,
namely, the inner Cauchy horizon and the black hole event horizon.
With the increasing of $\alpha_1$, the event horizon is shrinking.
When $\alpha_1=0$, the inner Cauchy horizon and the black hole
event horizon coincide and the solution corresponds to the extreme
solution.

\begin{figure}
\includegraphics[width=8.4cm]{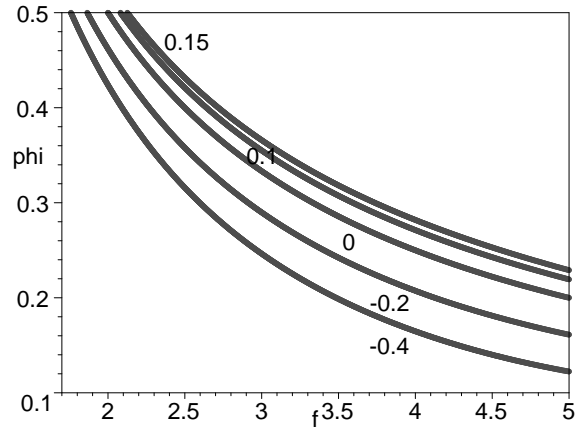}
\\
\caption{The evolution of the electric potential $\phi$ with
respect to the physical radius ${f}$ for different
$\alpha_1=-0.4,\ -0.2,\ 0, \, 0.1,\ 0.15$. It shows that $\phi$ is
regular in the spacetime except for $f=0$ (curvature singularity).
The potential $\phi$ is divergent at the curvature singularity and
asymptotically approaches zero in the infinity of space.}
\label{fig:fphi}
\end{figure}

In Fig.~\ref{fig:fphi}, we plot the evolution of the electric
potential $\phi$ with respect to the physical radius $f$ for
different $\alpha_1$. It shows that $\phi$ is regular in the
spacetime except for $f=0$ (curvature singularity). The potential
$\phi$ is divergent at $f=0$ and asymptotically approaches zero in
the infinity of space. This behavior is the same as the electric
potential in Reissner-Nordstrom solution.

\begin{figure}
\includegraphics[width=8.4cm]{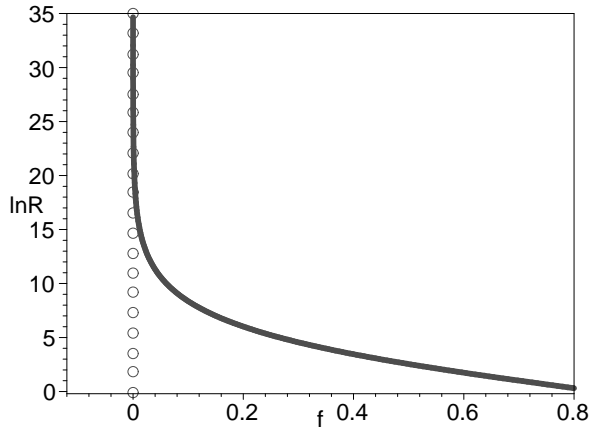}
\\
\caption{The evolution of the Ricci scalar $\ln R$ with respect to
the physical radius ${f}$ with $m=1,\ \alpha_0=\alpha_2=1,\ \
\alpha=3/2,\ \ \ \alpha_1=-0.4$. It is apparent $R$ is divergent
at $f=0$. This reveals that $f=0$ is indeed the curvature
singularity of spacetime.} \label{fig:fR}
\end{figure}

In order to show $f=0$ is the curvature singularity, as an
example, we plot the evolution of the Ricci scalar $R$ with
respect to the physical radius $f$ in Fig.~\ref{fig:fR} with
$m=1,\ \alpha_0=\alpha_2=1,\ \ \alpha=3/2,\ \ \ \alpha_1=-0.4$. It
is apparent $R$ is divergent at $f=0$. This reveals $f=0$ is
indeed the curvature singularity.

\begin{figure}
\includegraphics[width=8.4cm]{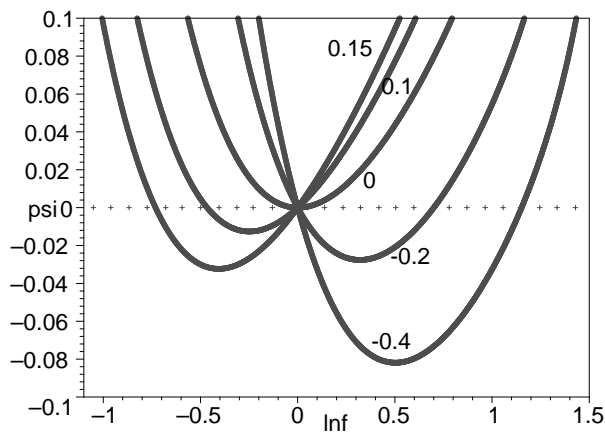}
\\
\caption{The evolution of the inverse of magnetic potential $\psi$
with respect to the physical radius $\ln{f}$ for different
$\alpha_1=-0.4,\ -0.2,\ 0, \, 0.1,\ 0.15$. It shows that the
magnetic potential $\psi^{-1}$ is divergent on the inner Cauchy
horizon and the outer black hole event horizon. On the curvature
singularity and the spatial infinity, it asymptotically approaches
zero.} \label{fig:fpsi}
\end{figure}

In Fig.~\ref{fig:fpsi}, we plot the evolution of the inverse of
magnetic potential $\psi$ with respect to the physical radius $
\ln f$ for different $\alpha_1$. It shows that the magnetic
potential $\psi^{-1}$ is divergent on both horizons while
asymptotically approaches zero in both the infinity of space and
the curvature singularity.

\begin{figure}
\includegraphics[width=8.4cm]{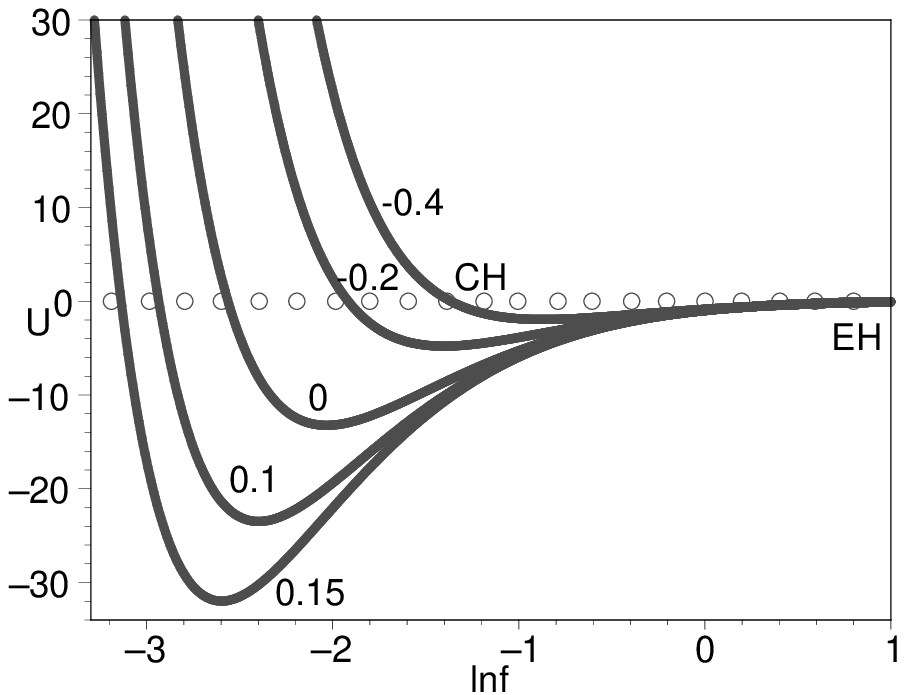}
\\
\caption{The evolution of $U$ with respect to the physical radius
$\ln{f}$ with values $m=1,\ \alpha_1=\alpha_2=1,\ \ \alpha=3/2$.
As for $\alpha_0$, we let $\alpha_0=0.15,\ 0.1,\ 0,\ -0.2,\ -0.4$,
respectively. Comparing with Fig.~\ref{fig:fu}, we find the black
hole event horizon is pushed to infinity in this case. We are left
with uniquely the inner Cauchy horizon. As an example, the
$\textrm{CH}$ and $\textrm{EH}$ are given for $\alpha_0=-0.4$. }
\label{fig:fu0}
\end{figure}

In Fig.~\ref{fig:fu0}, we plot the evolution of $U$ with respect
to the physical radius $ \ln f$ with values $m=1,\
\alpha_1=\alpha_2=1,\ \ \alpha=3/2$. As for $\alpha_0$, we let
$\alpha_0=0.15,\ 0.1,\ 0,\ -0.2,\ -0.4$, respectively. Comparing
with Fig.~\ref{fig:fu}, we find the black hole event horizon is
pushed to infinity in this case. We are left with only the inner
Cauchy horizon. Keep the constants ($\alpha_0,\ \alpha_1,\ \
\alpha_2,\ \ m,\ \ f_0$) to be fixed and verify  $\alpha$, we find
the figures are similar to Fig.~\ref{fig:fu} or
Fig.~\ref{fig:fu0}.

Finally, in order to understand the structure of horizons very
well, it would be very helpful to investigate the trajectories of
geodesic (free fall) paths in the spacetime. For simplicity, we
shall restrict ourselves to timelike and radial geodesics. The
equations of motion could be derived from the Lagrangian
\begin{eqnarray}
\mathscr{L}=\frac{1}{2}\left[U\dot{t}^2-\frac{1}{U}\dot{r}^2-f^2\dot{r}^2-f^2\sin^2\theta\dot{\varphi}^2\right]\;,
\end{eqnarray}
where the dot denotes the differentiation with respect to the
proper time $\tau$. They could also be derived from the geodesic
equation

\begin{eqnarray}
\frac{d^2X^{\mu}}{d\tau^2}+\Gamma^{\mu}_{\alpha\beta}\cdot\frac{dX^{\alpha}}{d\tau}\cdot\frac{dX^{\beta}}{d\tau}=0\;.
\end{eqnarray}
 The equations of motion are found to be

\begin{eqnarray}
\frac{dr}{d\tau}=-\sqrt{E^2-U}\;,
\end{eqnarray}
for proper time and
\begin{eqnarray}
\frac{dr}{dt}=-\frac{U}{E}\cdot\sqrt{E^2-U}\;,
\end{eqnarray}
for coordinate time $t$, respectively. Here $E$ is a constant.

We shall consider the trajectories of particles which start from
rest at some finite distance $r_0$ and fall towards the center.
The constant $E$ is related to the starting distance $r_0$ by

\begin{eqnarray}
E=\sqrt{U\mid_{r=r_0}}\;, \ \ (r=r_0\ \ \ \textrm{when}\ \
\dot{r}=0)
\end{eqnarray}
\begin{figure}
\includegraphics[width=8.4cm]{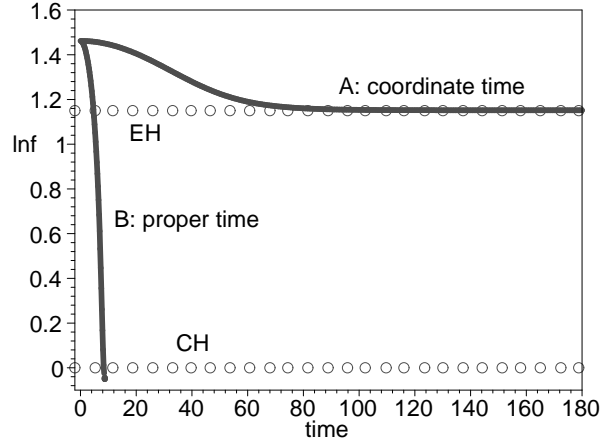}
\\
\caption{The evolution of coordinate time $t$ and proper time
$\tau$ along a timelike radial geodesics of a test particle,
starting at rest at $r_0=0.9\;$ or $\ln f_0=1.46$ and falling
towards the singularity.} \label{fig:ff}
\end{figure}

In Fig.~\ref{fig:ff}, we plot the evolution of coordinate time $t$
and proper time $\tau$ along the timelike radial geodesics. The
test particle starts at rest at $r_0=0.9\;$ or $\ln f_0=1.46$ and
falls towards the singularity. The same as the Figures (1-6), the
parameters are assumed with $f_0=1$, $m=1,\ \alpha_0=\alpha_2=1,\
\ \alpha=3/2$, $\alpha_1=-0.4$. The circled lines denote the black
hole event horizon (EH) and the Cauchy horizon (CH),
respectively(also shown in Fig.\ 1 and Fig.\ 3). Line A denotes
the coordinate time $t$. It shows that with respect to an observer
stationed at infinity, a particle describing a timelike trajectory
will take an infinite time to reach the black hole event horizon.
The behavior is in sharp contrast with that of proper time. Line B
denotes the evolution of proper time $\tau$. It shows that the
particle crosses the black hole event horizon and the Cauchy
horizon with finite proper time. And after crossing the Cauchy
horizon, the particle will arrive at some finite distance with
finite proper time.
\section{check of the solution with the Einstein equations }
In section II, we construct the static spherically symmetric
solution by imposing the symmetries of interest-rotational
symmetry and statistic-on the action principle rather than on the
field equations. Compared to the method of solving Einstein
equations, it is relatively simple, but also seems questionable.
The question is as follows. In imposing the symmetry before
carrying the variation of the action principle, one generally
loses field equations. So one may worry about that the solution
maybe do not satisfy the lost equations contained in the Einstein
equations. In this section, we shall check our solution with the
Einstein equations. To this end, we should start from the total
action of the theory which is given by

\begin{eqnarray}
S=\int
d^4x\sqrt{-g}\left[-{R}-\frac{c_1}{2}F_{\mu\nu}F^{\mu\nu}-c_2\left(\nabla_{\mu}A^{\mu}\right)^2+\lambda\left(A_{\mu}A^{\mu}+m^2\right)\right]\;.
\end{eqnarray}

In the first place, variation of the action with respect to
$\lambda$, we obtain the equation of motion for $\lambda$

\begin{eqnarray}\label{eq:check1}
A_{\mu}A^{\mu}+m^2=0\;.
\end{eqnarray}
Actually, it is the fixed-norm constraint on the aether field.

Secondly, variation of the action with respect to $A^{\mu}$ leads
to the equation of motion for aether field
\begin{eqnarray}\label{eq:check2}
c_1\nabla_{\nu}F^{\nu}_{\ \
\mu}+c_2\nabla_{\mu}\left(\nabla_{\nu}A^{\nu}\right)+\lambda
A_{\mu}=0\;.
\end{eqnarray}
This equation determines the dynamics of $A^{\mu}$. Finally,
variation of the action with respect to the metric gives the
Einstein equations

\begin{eqnarray}\label{eq:check3}
G_{\mu\nu}=T_{\mu\nu}\;.
\end{eqnarray}
We emphasize that the equation of fixed-norm constraint
Eq.~(\ref{eq:check1}) could be followed from the equation of
motion of aether field Eq.~(\ref{eq:check2}) and the Einstein
equations Eq.~(\ref{eq:check3}) in view of the fact that:
\begin{eqnarray}\label{eq:cons}
\nabla_{\nu}T^{\mu\nu}=0\;.
\end{eqnarray}
The energy-momentum tensor of the Einstein-aether field takes the
form \cite{picon:09}

\begin{eqnarray}
T_{\mu\nu}&=&c_1F_{\mu\alpha}F_{\nu}^{\
\alpha}+c_2g_{\mu\nu}\left[A^{\alpha}\nabla_{\alpha}\left(\nabla_{\beta}A^{\beta}\right)+\left(\nabla_{\alpha}A^{\alpha}\right)^2\right]-2c_2A_{(\mu}\nabla_{\nu)}
\left(\nabla_{\alpha}A^{\alpha}\right)\nonumber\\&&-\lambda
A_{\mu}A_{\nu}+\frac{1}{2}g_{\mu\nu}\left[-\frac{c_1}{2}F_{\alpha\beta}F^{\alpha\beta}-c_2\left(\nabla_{\alpha}A^{\alpha}\right)^2
-\lambda\left(A_{\alpha}A^{\alpha}+m^2\right)\right]\;.
\end{eqnarray}

Given the metric Eq.~(\ref{eq:line}) and the aether field
Eq.~(\ref{eq:A}), we find the equation of motion for $\lambda$:

\begin{eqnarray}\label{eq:lambda0}
-\frac{1}{U}\phi^2+\frac{K^2}{Uf^4}+m^2=0\;,
\end{eqnarray}
the equation of motion for $A_{\mu}$:
\begin{eqnarray}\label{eq:phi0}
\label{eq:phi} c_1Uf\phi^{''}+2c_1U\phi^{'}f^{'}+\lambda\phi
f=0\;,
\end{eqnarray}
\begin{eqnarray}\label{eq:K0}
c_2UfK^{''}-2c_2U K^{'}f^{'}+\lambda K f=0\;.
\end{eqnarray}
and the Einstein equations:
\begin{eqnarray}\label{eq:00E}
\left(1-2Uff^{''}-U^{'}ff^{'}-Uf^{'2}\right)\cdot\frac{1}{f^2}&=&\frac{1}{2}c_1\phi^{'2}-\frac{1}{2}c_2K^{'2}\frac{1}{f^4}+\lambda\left(\frac{K^2}{2Uf^4}-\frac{1}{2}m^2
-\frac{1}{2}\frac{\phi^{2}}{U}\right)\;,\\
\left(1-Uf^{'2}-U^{'}ff^{'}\right)\cdot\frac{1}{f^2}&=&\frac{1}{2}c_1\phi^{'2}-\frac{1}{2}c_2K^{'2}\frac{1}{f^4}+\lambda\left(-\frac{K^2}{2Uf^4}-\frac{1}{2}m^2
+\frac{1}{2}\frac{\phi^{2}}{U}\right)\;,\\
-\frac{1}{2}U^{''}-Uf^{''}\frac{1}{f}-U^{'}f^{'}\frac{1}{f}&=&-\frac{1}{2}c_1\phi^{'2}-\frac{1}{2}c_2K^{'2}\frac{1}{f^4}-\lambda\left(-\frac{K^2}{2Uf^4}+\frac{1}{2}m^2
-\frac{1}{2}\frac{\phi^{2}}{U}\right)\;,
\end{eqnarray}
which correspond to $G_0^0=T_0^0$, $G_1^1=T_1^1$ and $G_2^2=T_2^2$
, respectively. Now we have six equations of motion but five
variables, namely, $U,\ f,\ \phi,\ K,\ \lambda$. Therefore, among
the six equations, only five of them are independent. It is indeed
the case when we take into account the fact that the equation of
motion for $\lambda$ Eq.~(\ref{eq:check1}) follows from the
equation of motion of aether field Eq.~(\ref{eq:check2}) and the
Einstein equations Eq.~(\ref{eq:check3}). In practice, one could
show that Eq.~(\ref{eq:lambda0}) follows from Eqs.(56-60) by using
of Eq.~(\ref{eq:cons}).

One may ask whether the five equations of motion Eqs.~(9-13)
derived with the Euler-Lagrange method could be derived from above
six equations of motion Eqs.~(55-66). The answer is yes. In fact,
we have
\begin{eqnarray}
&&\textrm{Eq.}\left(9\right)\Longleftrightarrow\textrm{Eq.}\left(55\right)\;,\nonumber\\
&&\textrm{Eq.}\left(10\right)\Longleftrightarrow\textrm{Eq.}\left(56\right)\;,\nonumber\\
&&\textrm{Eq.}\left(11\right)\Longleftrightarrow\textrm{Eq.}\left(57\right)\;,\nonumber\\
&&\textrm{Eq.}\left(12\right)\Longleftrightarrow\textrm{Eq.}\left(58\right)-\textrm{Eq.}\left(59\right)\;,\nonumber\\
&&\textrm{Eq.}\left(13\right)\Longleftrightarrow\textrm{Eq.}\left(60\right)\;.\nonumber
\end{eqnarray}
Now we could understand that our solution satisfies all the
equations: the fixed-norm constraint equation, the equation of
motion of $A_{\mu}$ and the Einstein equations.

\section{conclusion and discussion}
In conclusion, a static spherically symmetric solution in the
Einstein-aether is obtained. Due to the complication of the
Einstein equations, we prefer to deal with the Euler-Lagrange
equations. This method is relatively simple and the same as the
Einstein equations in nature. By this way, an exact solution is
constructed. The solution is similar to the Reissner-Nordstrom
solution in that it has an inner Cauchy horizon and an outer black
hole event horizon. But a remarkable difference from the
Reissner-Nordstrom solution is that it is not asymptotically flat
in space. We find the solution asymptotically approaches a two
dimensional sphere. The resulting electric potential is regular in
the whole spacetime except for the curvature singularity. On the
other hand, the magnetic potential is divergent on both Cauchy
horizon and the outer event horizon.

 \acknowledgments

We sincerely thank the anonymous referee for the expert and
insightful comments which have significantly improved the paper.
We also thank Prof. Ted Jacobson for the very helpful discussions.
This work is supported by the National Science Foundation of China
under the Grant No. 10973014 and the 973 Project (No.
2010CB833004).

\newcommand\ARNPS[3]{~Ann. Rev. Nucl. Part. Sci.{\bf ~#1}, #2~ (#3)}
\newcommand\AL[3]{~Astron. Lett.{\bf ~#1}, #2~ (#3)}
\newcommand\AP[3]{~Astropart. Phys.{\bf ~#1}, #2~ (#3)}
\newcommand\AJ[3]{~Astron. J.{\bf ~#1}, #2~(#3)}
\newcommand\APJ[3]{~Astrophys. J.{\bf ~#1}, #2~ (#3)}
\newcommand\APJL[3]{~Astrophys. J. Lett. {\bf ~#1}, L#2~(#3)}
\newcommand\APJS[3]{~Astrophys. J. Suppl. Ser.{\bf ~#1}, #2~(#3)}
\newcommand\JHEP[3]{~JHEP.{\bf ~#1}, #2~(#3)}
\newcommand\JCAP[3]{~JCAP. {\bf ~#1}, #2~ (#3)}
\newcommand\LRR[3]{~Living Rev. Relativity. {\bf ~#1}, #2~ (#3)}
\newcommand\MNRAS[3]{~Mon. Not. R. Astron. Soc.{\bf ~#1}, #2~(#3)}
\newcommand\MNRASL[3]{~Mon. Not. R. Astron. Soc.{\bf ~#1}, L#2~(#3)}
\newcommand\NPB[3]{~Nucl. Phys. B{\bf ~#1}, #2~(#3)}
\newcommand\CQG[3]{~Class. Quant. Grav.{\bf ~#1}, #2~(#3)}
\newcommand\PLB[3]{~Phys. Lett. B{\bf ~#1}, #2~(#3)}
\newcommand\PRL[3]{~Phys. Rev. Lett.{\bf ~#1}, #2~(#3)}
\newcommand\PR[3]{~Phys. Rep.{\bf ~#1}, #2~(#3)}
\newcommand\PRD[3]{~Phys. Rev. D{\bf ~#1}, #2~(#3)}
\newcommand\RMP[3]{~Rev. Mod. Phys.{\bf ~#1}, #2~(#3)}
\newcommand\SJNP[3]{~Sov. J. Nucl. Phys.{\bf ~#1}, #2~(#3)}
\newcommand\ZPC[3]{~Z. Phys. C{\bf ~#1}, #2~(#3)}
 \newcommand\IJGMP[3]{~Int. J. Geom. Meth. Mod. Phys.{\bf ~#1}, #2~(#3)}
  \newcommand\GRG[3]{~Gen. Rel. Grav.{\bf ~#1}, #2~(#3)}

\end{document}